\documentclass[10pt,journal,compsoc]{IEEEtran}
\usepackage{moreverb,url}

\usepackage{subfigure}
\usepackage{pgf}
\usepackage{tikz}
\usepackage{tabularx}
\newcolumntype{P}[1]{>{\centering\arraybackslash}b{#1}}
\usetikzlibrary{arrows,automata, positioning, shapes, decorations.pathreplacing}
\usepackage{array}
\usepackage{pbox}

\usepackage{amsmath}
\usepackage[ruled]{algorithm2e}
\usepackage{algcompatible}
\usepackage{floatrow}
\usepackage{graphicx}
\graphicspath{ {Figures/} }

\newcommand\mrp{\mathrel{\overset{\makebox[0pt]{\mbox{\normalfont\tiny\sffamily m.r.p.}}}{\equiv}}}
\newcommand\mnp{\mathrel{\overset{\makebox[0pt]{\mbox{\normalfont\tiny\sffamily m.n.p.}}}{\equiv}}}

\ifCLASSOPTIONcompsoc
  \usepackage[nocompress]{cite}
\else
  \usepackage{cite}
\fi

\ifCLASSINFOpdf
\else
\fi

\begin{document}

\title{BB-Graph: A Subgraph Isomorphism Algorithm for Efficiently Querying Big Graph Databases}

\author{Merve~Asiler
        and~Adnan~Yaz{\i}c{\i},~\IEEEmembership{Senior Member,~IEEE}
\IEEEcompsocitemizethanks{\IEEEcompsocthanksitem M. Asiler and A. Yaz{\i}c{\i} are with the Department
of Computer Engineering, Middle East Technical University, Ankara, 06800
Turkey. \protect\\
E-mail: \{asiler, yazici\}@ceng.metu.edu.tr.
\IEEEcompsocthanksitem A. Yazici is currently with the Nazarbayev University, Department of
Computer Science at School of Science and Technology, Astana, Kazakhstan. \protect\\
E-mail: adnan.yazici@nu.edu.kz.}% <-this % stops an unwanted space
}

\IEEEtitleabstractindextext{%
\begin{abstract}
The big graph database model provides strong modeling for complex applications and efficient querying. However, it is still a big challenge to find all exact matches of a query graph in a big graph database, which is known as the subgraph isomorphism problem. The current subgraph isomorphism approaches are built on Ullmann's idea of focusing on the strategy of pruning out the irrelevant candidates. Nevertheless, the existing pruning techniques need much more improvement to efficiently handle complex queries. Moreover, many of those existing algorithms need large indices requiring extra memory consumption. Motivated by these, we introduce a new subgraph isomorphism algorithm, named as BB-Graph, for querying big graph databases efficiently without requiring a large data structure to be stored in main memory. We test and compare our proposed BB-Graph algorithm with two popular existing approaches, GraphQL and Cypher. Our experiments are done on three different data sets; (1) a very big graph database of a real-life population database, (2) a graph database of a simulated bank database, and (3) the publicly available World Cup big graph database. We show that our solution performs better than those algorithms mentioned here for most of the query types experimented on these big databases.
\end{abstract}

}

\maketitle

\IEEEdisplaynontitleabstractindextext

\IEEEpeerreviewmaketitle

% INTRODUCTION

\IEEEraisesectionheading{
\section{INTRODUCTION}
\label{sec:introduction}}

\IEEEPARstart{I}{n} the last decade, the big graph database models have widespreaded in a large variety of application areas such as communications, logistics, Web/ISV, network management, social networks, mobile communication applications, data center management, bioinformatics, etc. The graph database model has recently been preferred to the other database models because it better fits into structure of various complex queries on big data providing higher performance than the others, especially RDBMs, for most complex cases \cite{vojtechkolomicenko2013}, \cite{batra2012comparative}, \cite{vicknair2010comparison}, \cite{bitninecompany2014}.

\begin{figure} [!h]
	\centering
	\resizebox{0.8\textwidth}{!}{
	\subfigure[query graph: $Q$]{
		\begin{tikzpicture}
		[->,>=stealth',shorten >=1pt,auto,node distance=1.5cm, semithick]
		\tikzset{vertex/.style = {shape=circle,draw,minimum size=1.5em}}
		\tikzset{edge/.style = {->}}

		\node[vertex, label=right:$u_{2}$]					(2)	{$B$};
		\node[vertex, label=$u_{0}$, above left of=2]	(0)	{$A$};
		\node[vertex, label=$u_{1}$, above right of=2]	(1)	{$B$};
		\node[vertex, label=right:$u_{3}$, below of=2]		(3)	{$D$};
		
		\path
		(0)	edge (2)
		(1)	edge (0)
			edge (2)
		(2)	edge (3);
	
		\end{tikzpicture}}
	\hfill
	\subfigure[graph database: $G$]{
		\begin{tikzpicture}
		[->,>=stealth',shorten >=1pt,auto,node distance=1.5cm, semithick]
		\tikzset{vertex/.style = {shape=circle,draw,minimum size=1.5em}}
		\tikzset{edge/.style = {->}}
		
		\node[vertex, label=$\nu_{0}$]									(0)	{$B$};
		\node[vertex, label=$\nu_{3}$, below left of=0]					(3)	{$B$};
		\node[vertex, label=$\nu_{1}$, above left of=3]					(1)	{$A$};
		\node[vertex, label=below:$\nu_{2}$, below left of=3] at (-1.5,-0.5)	(2)	{$B$};
		\node[vertex, label=left:$\nu_{4}$, below of=3]						(4)	{$D$};
		\node[vertex, label=below:$\nu_{5}$, below of=0]						(5)	{$C$};
		\node[vertex, label=$\nu_{7}$, below right of=0]					(7)	{$B$};
		\node[vertex, label=$\nu_{6}$, above right of=7]					(6)	{$A$};
		\node[vertex, label=right:$\nu_{8}$, below of=7]						(8)	{$D$};
		\node[vertex, label=below:$\nu_{9}$, below right of=7] at (1.5, -0.5)	(9)	{$A$};
		
		\path
		(0)	edge (1)
			edge (3)
			edge (5)
		(1)	edge (3)
		(2)	edge (1)
			edge (3)
		(3)	edge (4)
		(6)	edge (0)
		(7)	edge (0)
			edge (6)
			edge (8)
			edge (9);
		
		\end{tikzpicture}}}
	\caption{An example query and a graph database with unlabeled relationships}
	\label{fig:sample_query_database}
\end{figure}
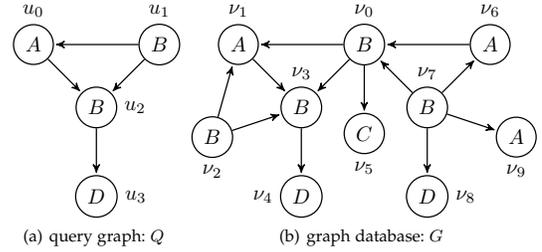

On the other hand, the subgraph isomorphism problem is one of the most frequently encountered challenges in big graph database applications. Subgraph isomorphism can be defined as follows: Given a query graph $Q$ and a database graph $G$, find all matching instances of $Q$ in $G$. Figure \ref{fig:sample_query_database} illustrates this, where there exist two instances of $Q$ in $G$, one is the subgraph consisting of the nodes $\nu_{0}$, $\nu_{1}$, $\nu_{3}$, $\nu_{4}$ and the relationships $_{\nu_{0}}e_{\nu_{1}}$, $_{\nu_{0}}e_{\nu_{3}}$, $_{\nu_{1}}e_{\nu_{3}}$, $_{\nu_{3}}e_{\nu_{4}}$ and the other one is the subgraph consisting of the nodes $\nu_{1}$, $\nu_{2}$, $\nu_{3}$, $\nu_{4}$ and the relationships $_{\nu_{1}}e_{\nu_{3}}$, $_{\nu_{2}}e_{\nu_{1}}$, $_{\nu_{2}}e_{\nu_{3}}$, $_{\nu_{3}}e_{\nu_{4}}$. The subgraph isomorphism problem is known as an NP-hard problem \cite{abboud2016subtree}. In almost all big graph database applications, there frequently occur queries directly or indirectly related to subgraph isomorphism problem; therefore, it is important to find an efficient solution for the subgraph isomorphism problem to handle complex queries on big databases.

Most of the subgraph isomorphism algorithms in the literature are based on one of the following two types of strategies: Feature indexing by using \textit{filtering-and-verification} technique or candidate node checking by using \textit{branch-and-bound} technique. Algorithms of the first type create an index of small graphs (features) and filter the database pieces which do not have some (or all) of the features included by the query graph. They mark the ones which succesfully passed the filtering to include an exact match(es) of the query, and try to verify them with respect to whether they really do or not. GraphGrep \cite{giugno2002graphgrep}, GIndex \cite{yan2004graph}, Labeled Walk Index (LWI) \cite{srinivasa2005lwi}, Closure-Tree \cite{he2006closure}, Graph Decomposition Indexing \cite{williams2007graph}, TreePi \cite{zhang2007treepi}, TreeDelta \cite{zhao2007graph}  are examples to this type of algorithms. Although those algorithms are good for decreasing the number of candidate data sets, they cannot deduce all the isomorphisms of a query graph. Moreover, they are applicable only for the databases consisting of many disconnected graph pieces. On the other hand, the algorithms of the second type aim to find all the exact matches of a query graph without making an issue that the database is connected or disconnected. Starting off from Ullmann's idea \cite{ullmann1976algorithm}, they extract all the candidates for each node in query and, incrementally, they match a query node with one of its candidates. In that way, they try to reach an exact match of a query by branching the previous matches. VF2 \cite{cordella2004sub}, VF3 \cite{carletti2017introducing}, QuickSI \cite{shang2008taming}, GADDI \cite{zhang2009gaddi}, GraphQL \cite{he2008graphs}, SPath \cite{zhao2010graph} belong to this type of algorithms. The mentioned algorithms are good in terms of finding all of the exact matches of a query graph; nevertheless, since they search the candidates globally along the database, most of their tryouts to find a relationship between the nodes are redundant due to the fact that the nodes are irrelevant. Therefore, they have poor computational performance. Additionally, in order to prune the nodes which cannot be a candidate, some of those algorithms need large data structures, which results in an extra amount of memory consumption.  

Motivated by all mentioned above, we introduce a new \textit{branch-and-bound} algorithm combined with backtracking strategy for graph isomorphism problem, called BB-Graph. Our proposed algorithm removes the drawbacks mentioned above by conducting the search of candidates for other query nodes locally in the close neighbourhood of the firstly-matched database node. Therefore, we can summarize the main contributions of our study as follows:

\begin{enumerate}
	\item A new subgraph isomorphism algorithm, BB-Graph, is introduced. BB-Graph follows a more efficient search strategy than the other existing algorithms while matching graph elements (nodes and relationships) by using \textit{branch-and-bound} technique combined with backtracking: Initially, candidates for the starting query node are selected from all across the graph database using some node properties. Then for each candidate, an isomorphism region is potentially created by matching the starting node with that candidate. Next, the candidates for the other query nodes are selected locally by traversing every neighbour node and relationship using branch-and-bound method. Later, our algorithm backtracks to check possible relationships and node combinations not yet handled.

	\item BB-Graph uses the built-in data structures. Thus, it does not consume extra memory for large indexing or data storing.
		
	\item While many of the other algorithms were experimented only on undirected graphs with hundreds of thousands of nodes by reading the input from a BLOB file, BB-Graph has been tested and evaluated on a very big connected directed graph with almost 70 millions of nodes in addition to two other directed graphs, one with 100 thousands of nodes and the other one with 45 thousands of nodes.
		
	\item A number of experiments done for comparing our algorithm (BB-Graph) with Cypher and GraphQL show that BB-Graph has a better computational performance than the existing algorithms for most of query types.
	
	\item The effect of a change in node matching order is illustrated with the experiments done during this study. The results show that developing an effective node matching order is definitely a considering issue to increase the performance of a subgraph isomorphism algorithm and our study can definitely guide those who want to work in this topic.
\end{enumerate}

Organization of the rest of this paper is as follows: Section \ref{sec:literature} gives the necessary background and related work. Section \ref{sec:algorithm} introduces our BB-Graph approach and the main algorithms along with the pseudocodes of our proposed solutions. Section \ref{sec:experimental.work} shows the experimental results and comparison of BB-Graph with Cypher and GraphQL. Lastly, the conclusion and feature work are given in Section \ref{sec:conclusion}.

% LITERATURE SEARCH

\section{Background and Related Work}
\label{sec:literature}

\begin{figure} [!h]
	\centering
	\resizebox{1\textwidth}{!}{
		
	\includegraphics{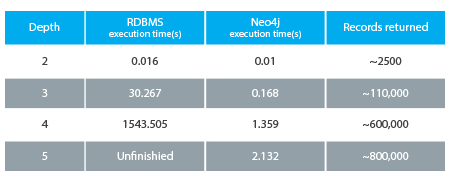}}
	
	\caption{Comparison results for running time performaces of searching a recursive query with difference depths in RDBMs and Neo4j \cite{bitninecompany2014}} 
	\label{fig:recursive_query}
\end{figure}

Recently, graph databases have been popular since they can efficiently handle some database operations required for complex queries on big databases. Many existing studies comparing the graph databases and relational databases show that graph databases perform much better than RDBMs with respect to many aspects \cite{vojtechkolomicenko2013}, \cite{batra2012comparative}, \cite{vicknair2010comparison}, \cite{bitninecompany2014}, \cite{lange2001comparing}, \cite{wycislik2014performance}, \cite{miller2013graph}, \cite{nayak2013type}. According to those studies, generally, the join operations on relational tables cause high costs in querying RDBMs. On the other hand, according to their experimental results previously done in the literature, such a problem disappears in graph databases, mostly because of directly traversing the nodes and relationships. In Figure \ref{fig:recursive_query}, the results of an experiment done with recursive queries that require many join operations are given \cite{bitninecompany2014}. Experimental results of the comparison between relational databases and graph databases show that graph databases have quite higher performance than RDBMs for such complex queries \cite{kuccukkecceci2016graph}. Moreover, the previous studies also show that insert-delete-update operations on graph databases can be done much more easily and efficiently compared to relational databases.

\subsection{Background}
\label{subsec:background}

In this paper, the notation $G$ : $(V, E)$ represents the graph $G$ with the vertex set $V$ and edge set $E$. For vertex $\nu$; $L_{\nu}$ denotes the label set of $\nu$. Similarly, for edge $e$, $L_{e}$ denotes the label set of $e$. The notation $_{u}e_{\nu}$ is used for the edge $e$ starting from vertex $u$ (outgoing w.r.t. $u$) and ending at vertex $\nu$ (incoming w.r.t. $\nu$) when we need to state the end points of $e$. Otherwise, we just denote it as $e$. 

For two graphs $G_{1}:(V_{1}, E_{1})$ and $G_{2}:(V_{2}, E_{2})$, $G_{1}$ is said to be \textit{subisomorphic} to $G_{2}$, if there is a one-to-one and onto function $f: V_{1} \longrightarrow V_{2}$ such that; $\forall \nu \in V_{1}$ $L_{\nu} \subseteq L_{f(\nu)}$ and for any $_{u}e_{\nu}$ $\in$ $E_{1}$ $_{f(u)}e_{f(\nu)}$ $\in$ $E_{2}$ satisfying that $L_{_{u}e_{\nu}} \subseteq L_{_{f(u)}e_{f(\nu)}}$. Given a query graph $Q$ and a data graph $G$, the problem of finding all distinct subisomorphisms of $Q$ in $G$ is called as \textit{Subgraph Isomorphism Problem} \label{def:ProblemDefinition}.

For two graphs $G_{1}:(V_{1}, E_{1})$ and $G_{2}:(V_{2}, E_{2})$, where $G_{1}$ is subisomorphic to $G_{2}$ under some isomorphism $f$, let $u \in V_{1}$ be matched with $u':=f(u) \in V_{2}$. Since each edge $e$ adjacent to $u$ must be matched with some edge $e'$ adjacent to $u'$, it follows that the degree of $u$ must be less than or equal to the degree of $u'$. Moreover, since the matched edges must have the same label and direction, for each group of edges having the same label and direction degree of $u$ must be also less than or equal to the degree of $u'$. As a result, the matching vertices in a subisomorphism must satisfy both label and degree rules deduced from the definition. In this paper, we call this as \textit{matching node principal} for the vertices to be matched \label{def:matchingnodeprincipal}. We denote $u \mnp u'$ when matching node principle holds for $u$ and $u'$.  Similarly, if $e$ is matched with $e'$  under $f$, then it means $L_{e} \subseteq L_{e'}$ and direction of $e$ with respect to $u$ must be the same with direction of $e'$ with respect to $u'$. We call this as \textit{matching relationship principal} for the relationships to be matched and denote $e \mrp e'$\label{def:matchingrelationshipprincipal}.

Throughout the paper, we use the word \textit{relationship} instead of the \textit{edge} concept and the word \textit{node} instead of \textit{vertex} concept to make the contextual narrating compatible with Neo4j environment. Moreover, we denote the matching of query node $u$ with database node $u'$ by $<u, u'>$; similarly, the matching of query relationship $r$ with the database relationship $r'$ by $<r, r'>$.

\subsection{Ullmann Algorithm And Its Derivations}
\label{subsec:UllmannDerivations}

Ullmann Algorithm \cite{ullmann1976algorithm} is the first search method that is developed to find isomorphic patterns of query graphs in large graphs. It basically consists of 4 main steps:
\begin{itemize}
	\item Filtering candidates for each query node, 
	\item Selecting a candidate for each query node, trying to match the node with that candidate together by matching the corresponding relationships between the currently processed node and previously matched nodes,
	\item Replacing the selected candidate with another one if the current match does not work, and
	\item Backtracking in order to try other candidates and to find more isomorphisms.
\end{itemize}

\begin{figure} [!h]
	\centering
	\resizebox{0.45\textwidth}{!}{
		\subfigure[graph A]{
			\begin{tikzpicture}
			[->,>=stealth',shorten >=1pt,auto,node distance=2.5cm, semithick]
			\tikzset{vertex/.style = {shape=circle,draw,minimum size=1.5em}}
			\tikzset{edge/.style = {->}}
			
			% graph 1: nodes
			\node[vertex, label=$u_{0}$]		(1) at (2,2)	{$A$};
			\node[vertex, label=below:$u_{2}$]	(2) at (2,0)	{$B$};
			\node[vertex, label=right:$u_{4}$]	(4) at (2,-2)	{$B$};
			\node[vertex, label=$u_{3}$]		(3) at (4,0)	{$C$};
			\node[vertex, label=$u_{1}$]		(5) at (0,0)	{$D$};
			
			% graph 1: edges
			\path
			(1) edge [color = orange] (2)
			edge [color = red] (3)
			(3) edge (4)
			edge [color = purple] (2)
			(5) edge [color = green] (2)
			edge [color = blue](1)
			edge (4);
			
			\end{tikzpicture}}}
	\hfill
	\resizebox{0.45\textwidth}{!}{
		\subfigure[graph B]{
			\begin{tikzpicture}
			[->,>=stealth',shorten >=1pt,auto,node distance=2.5cm, semithick]
			\tikzset{vertex/.style = {shape=circle,draw,minimum size=1.5em}}
			\tikzset{edge/.style = {->}}
			
			% graph 2: nodes
			\node[vertex, label=$u'_{0}$]		(6) at (10,2)	{$A$};
			\node[vertex, label=$u'_{1}$]		(7) at (8,0)	{$D$};
			\node[vertex, label=below:$u'_{2}$]	(8) at (10,0)	{$B$};
			\node[vertex, label=$u'_{3}$]		(9) at (12,0)	{$C$};
			
			% graph 2: edges
			\path
			(6) edge [color = orange] (8)
			edge [color = red, dashed] node {?} (9)
			(7) edge [color = blue] (6)
			edge [color = green] (8)
			(9) edge [color = purple, dashed] node {?} (8);
			
			\end{tikzpicture}}}
	\caption{\textsc{IsJoinable} procedure of $u_{3}$ and $u'_{3}$ while matching the graphs A and B} 
	\label{fig:isjoinable}
\end{figure}
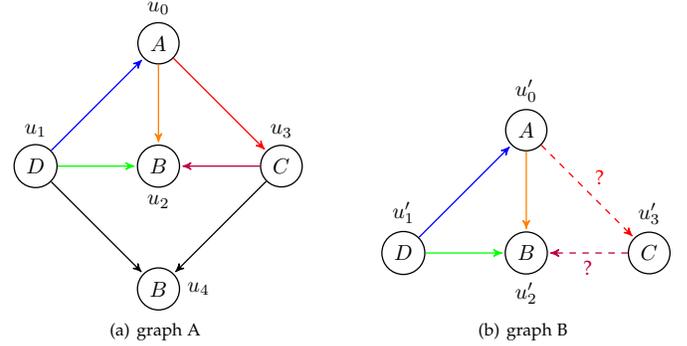

To filter the candidates for each query node, The Ullmann Algorithm checks for the matching node principal (described in Section \ref{def:matchingnodeprincipal}). After the filtering step, it starts with matching of the nodes in a recursive manner in order of the nodes given in the input. Matching two nodes $u$ and $u'$, it checks for each relationship between $u$ and some previously matched query node $\nu$, if there is a corresponding relationship between $u'$ and $\nu'$ which is the database node matched with $\nu$. In other words, to be intermateable, $u$ and $u'$ must satisfy that for each relationship $_{u}r_{\nu}$ (or $_{\nu}r_{u}$) where $<\nu, \nu'> \in V_{matched}$, there is a corresponding relationship $_ {u'}r'_{\nu'}$ (or $_{\nu'}r'_{u'}$, respectively) where $r \mrp r'$. In the study of Lee et al.\cite{lee2012depth}, the procedure to check the existence of this condition is called as \textsc{IsJoinable}. In Figure \ref{fig:isjoinable}, \textsc{IsJoinable} operation for the match of $u_{3}$ in graph A with $u'_{3}$ in graph B is illustrated. Each $u_{i}$ is matched with $u'_{i}$ where $0 \leq i \leq 2 $ and the relationships $_{u_{1}}e_{u_{0}}$, $_{u_{0}}e_{u_{2}}$, and $_{u_{1}}e_{u_{2}}$ are matched with $_{u'_{1}}e_{u'_{0}}$, $_{u'_{0}}e_{u'_{2}}$, and $_{u'_{1}}e_{u'_{2}}$, respectively. When the matching turn comes to $u_{3}$ and $u'_{3}$, \textsc{isJoinable} procedure searchs candidates for the relationships between $u_{3}$ and previously matched nodes $u_{0}$ and $u_{2}$ (the relationship between $u_{3}$ and $u_{4}$ is not considered since $u_{4}$ is not matched yet), so it checks if there is any corresponding relationship between $u'_{0}$ and $u'_{3}$ and between $u'_{2}$ and $u'_{3}$ which satisfy the necessary matching relationship principals, respectively.

In the matching phase of query node $u$ with database node $u'$, if there does not exist any problem upto the end of \textsc{IsJoinable} procedure, $<u,u'>$ is added into the list of matched nodes. Then, isomorphism search continues by picking the next not-yet-matched query node in order to match that with one of its candidates. In the case that \textsc{IsJoinable} procedure returns false for a query node and its matching node, the current match is cancelled and the next candidate is selected this time. In either case, when all the matched nodes result in an exact match of the query graph or one of the node matchings results in a failure, the algorithm backtracks to try other candidates.

The Ullmann Algorithm is reasonably efficient for finding all isomorphic patterns of the query graph in the large database graph and handling all possible node and relationship matchings through backtracking. Although it is a robust algorithm, performance of The Ullmann Algorithm can be increased with well-thought matching order strategies and effective pruning rules and some short cuts. Thereby, the following 5 algorithms are derived from Ullmann's Algorithm as we briefly discuss each one below. 

\textit{VF2} \cite{cordella2004sub} is one of those algorithms derived from Ullmann's Algorithm. It matches the nodes in an increasing order of number of their labels regarding a specified query graph by following their strategy. It selects the next query node from the set of nodes which are connected to at least one of the previously matched query nodes with a relationship. In this way, it can eliminate more candidates during \textsc{IsJoinable} stage. Moreover, VF2 algorithm refines the candidates before passing to \textsc{IsJoinable} step by comparing degrees of already matched and not-yet-matched neighbour nodes. It divides the not-yet-matched adjacent nodes into two, as the ones in first-degree-neighbourhood of the already matched nodes and the ones not in that area. Then it makes degree comparison separately for both of two sets adjacent to query node $u$ and their correspondents that are adjacent to candidate of $u$.	

Another algorithm derived from Ullmann's is \textit{QuickSI} \cite{shang2008taming}. The key aspect of this method is that it defines a data structure named as QI-Sequence which provides efficient pruning; and thus, resulting in low-cost processing. QI-Sequence is a minimum spanning tree created based on edges weighted according to the number of each node label and the number of each $<$start node label - relationship type - end node label$>$ triple in database graph. Starting to match from the relationships and nodes with low frequency, there occurs less possibilities to test and in this way QuickSI is able to decrease the number of recursive calls. Also, QuickSI uses QI-Sequence in indexing of the features in database graphs and takes advantage of its tree structure to prune candidate graphs. As its pruning strategy, QuickSI applies \textsc{IsJoinable} procedure over a query node $u$ by beginning the checks from the relationship between $u$ and its parent node in QI-Sequence (in case that $u$ is not the root node).

The third algorithm is \textit{GADDI} \cite{zhang2009gaddi}, introduced by Zhang et al., and is based on how to refine the candidate nodes by using a distance based indexing. The main purpose is to eliminate the cadidate nodes by examining their $k$-neighbourhood in case where there are not as many specific fragments as there are in the $k$-neighbourhood of query node. Selecting the fragments that are used in neighborhood comparison, the discriminative ones which occur with different frequencies in common $k$-neighbourhood of sample pairs of database node are picked. GADDI uses 3 different pruning rule to refine the candidates: For a query node $u$ and a candidate node $u'$, firstly for each node $\nu$ in $k$-neighborhood of $u$, it tries to find a candidate $\nu'$ in $k$-neighbourhood of $u'$ by comparing labels. Secondly, for the common $k$-neighbourhood of each $(u, \nu)$ pair, it counts the number of discriminative fragments in this area and prunes out $u'$ if there are less number of occurences of a specific fragment in the corresponding region obtained by $(u', \nu')$. Thirdly, for each $\nu$ it compares the length of shortest path, say $t$, between $u$ and $\nu$ with the one, $t'$, between $u'$ and $\nu'$ and $u'$ is eliminated in the case $t < t'$ for at least one $\nu$. Furthermore, GADDI applies these pruning rules in reverse manner for each candidate $\nu'$ in neighbourhood of $u'$. Lastly, as matching order, GADDI selects the first node randomly, the rest are selected by depth first search.

\textit{GraphQL} \cite{he2008graphs}, a popular algorithm for subgraph isomorphism problem, focuses on neigbourhood relations to filter candidates for a query node. If a query node $u$ can be matched with a database node $u'$, then for each query node $u_{k}$ in $k$-neighbourhood of $u$, there must be a candidate node $u_{k}'$ in $k$-neighbourhood of $u'$. Thus, GraphQL uses this fact to prune out false candidates of a query node, by scanning their $k$-neighbourhood upto a refinement level $l$, incrementally for each $k$ where $1 \leq k \leq l$. Also, GraphQL follows an optimized node matching strategy by selecting the query node which is estimated to decrease the cost at each intermediate step and adjacent to set of already matched nodes. GraphQL compares neighbourhoods of query nodes with their candidates' neighbourhoods and tries to find a semi-perfect bipartite matching between the nodes in corresponding neighbouhoods but our experiments show that this is already an exhaustive computation even when the refinement level is set to one and not very effective for reducing the candidate set size of some query nodes.

Lastly, as a candidate path matching version of the Ullmann's method, \textit{SPath} algorithm \cite{zhao2010graph} handles candidate paths. It actually does nothing but matches more than one node on a linear sequence at a recursive call by applying \textsc{IsJoinable} procedure for each node on the path. SPath algorithm filters the candidate vertices by checking the number of each node label in their $k$-neighbourhood, where $k$ is a parameter for the radius of neighbourhood. It applies a rule which is as follows: For each node label $L$; total number of occurences of $L$ in the neighbourhood upto $k^{th}$ level of query node $u$ must be less than or equal to the total number of occurences of $L$ in the neighbourhood upto the $k^{th}$ level of database node $u'$ where $u'$ is a candidate for $u$. While matching the paths, the algorithm follows a decreasing order of path selectivity defined as a metric based on size of candidate node sets.

The algorithms summarized here remain incapable of showing satisfying performance in some complex query cases when data is very big. Each algorithm has some drawbacks: The pruning techniques of VF2 are not powerful enough, and the matching order it follows is effective only when database graph has similar node label statistics with query graph. QuickSI has to go around the whole database to deduce the information about label and $<$label-relationship type-label$>$ triple count used in edge weighting; moreover, it should keep up-to-date data. Therefore it additionally needs B+-Tree index structures only for this purpose. GADDI creates a large index that keeps the  number of discriminative fragments in the intersected $k$-neighbourhood of each node pair in the whole database, which requires a very exhaustive pre-computation. Futhermore, the pruning rules of GADDI are not effective, and quite time-consuming. On the other hand, GraphQL compares neighbourhoods of query nodes with their candidates' neighbourhoods and tries to find a semi-perfect bipartite matching between the nodes in corresponding neighbouhoods. Nevertheless, our experiments show that this is already an exhaustive computation even when the refinement level is set to one and not very effective for reducing the candidate set size of some query nodes. SPath needs a data structure including the number of each label with shortest distance $i$ from $u'$ for each database node $u'$ and for each $i$ from 1 to $k$; which requires a long pre-computation time and large storage. Also, the experiments in \cite{lee2012depth} show that the ordering based on path selectivity does not provide a good performance for searching the graph in database.

In their study, Lee et al. \cite{lee2012depth} compare five subgraph isomorphism algorithms; VF2, QuickSI, GADDI, GraphQL, and SPath on some real-world data sets on iGraph framework \cite{han2010igraph}. It has been shown that these algorithms only work on undirected graphs consisting of one or many pieces by testing with subgraph, clique and path queries. The experiments show that there is not a satisfying algorithm which works for all types of database queries especially complex queries on directed big graphs efficiently. For instance, while QuickSI shows a good performance in many cases, it fails to return the answer in a reasonable time for NASA dataset described in \cite{lee2012depth}. According to the experimental results that they have obtained, GraphQL is the only algorithm succeeding to respond in a reasonable time for all tests on that dataset. They state that these start-of-the-art algorithms perform poorly because of their ineffective matching order and the trade-off between efficiency and overhead of their pruning methods. 

Although each of the five algorithms, VF2, QuickSI, GADDI, GraphQL, and SPath, has its own defect, there exists a common point that causes all of these algorithms to perform poorly. The pruning rules that they use are generally effective for eliminating the nodes that can never be a representative of the query node for which it is selected as a candidate in a real exact match. However, since they do not apply the prunings by regarding each isomorphism as an independent one, the nodes belonging to different isomorphisms can not be discriminated until their relationships are checked in matching phase. Therefore, the database nodes which are candidates for different query vertices and members of distinct isomorphisms seem available for taking place in the same subgraph isomorphism at first sight. Hence, the main time consumed occurs at this point while searching for an actually non-existing reasonable connection between those irrelevant nodes. In order to remove such cases, after the start node candidates are taken from all across the database, candidate nodes for the rest of the query vertices should be selected depending on the start node match. In other words, each starting node candidate potentially creates a distinct isomorphism region; therefore, the representatives for the other query nodes should be chosen from the close neighbourhood of the starting node through a \textit{local} region scanning instead of a \textit{global} one for each distinct exact match.

% ALGORITHM

\section{BB-GRAPH ALGORITHM}
\label{sec:algorithm}

Our proposed algorithm, BB-Graph, first chooses a query node to start matching, which we call it the \textit{starting node}. Next, BB-Graph filters the database nodes to find candidates for the starting node. It does this filtering according to the \textit{matching node principal} previously mentioned above. After getting the candidate nodes, BB-Graph puts them into a list to match the starting node with one candidate at a time and search for query graph isomorphisms rooting from that match. In order to achieve this, the whole candidate list is walked through a loop such that at each time of iteration the next candidate node is picked from the list, matched with the starting node and the recursive isomorphism search begins from that point, as shown in Algorithm \ref{alg:BB-Graph}, lines 1-12.

\begin{algorithm}[ht]
	\SetKwInput{Input}{Input}
	\SetKwInput{Output}{Output}
	\LinesNumbered
	\DontPrintSemicolon
	\BlankLine
	\Input	{$\it{Q:}$ Query graph with $n$ vertices\;\qquad \quad
			$\it{G:}$ Database graph}
	\Output{$\it{M:}$ Set of all exact matches of $Q$ in $G$}
	
	\Begin
	{
		$M := \emptyset$ \;
		Choose the first node in the input as $u_{start}$ \;
		$C_{u_{start}} := \{u' \; | \; u' \in V_{G} \text{ and } u' \mnp u_{start} \}$ \;
		\ForEach{$u' \in C_{u_{start}}$}
		{
			$V_{matched} := \emptyset, E_{matched} := \emptyset, S := \emptyset$ \tcp{Reset the temporary storage}
			Push $<u_{start}, u'>$ into $S$ \;
			Add $<u_{start}, u'>$ into $V_{matched}$ \;
			\textsc{Search()}
		}
		return $M$
	}
	\textbf{void} \textsc{Search()} \;
	\Begin
	{
		\If{$S \neq \emptyset$}
		{
			$<u, u'> :=$ Pop $S$  \;
			\If{$u$ \upshape{has non-matched adjacent relationship}}
			{
				\textsc{BranchNodes}($<u, u'>$)
			}
			\Else
			{
				\textsc{Search()}
			} 
		}
		\Else
		{
			Add the tuple $(V_{matched}, E_{matched})$ into $M$ \tcp{An exact match found}
		}
		return
	}
	\caption{\textsc{BB-Graph Search Algorithm}}
	\label{alg:BB-Graph}
\end{algorithm}

In isomorphism search, BB-Graph uses the previously obtained node matches. Let $V_{matched}$ be the set of previously matched couples of nodes. Similarly, let $E_{matched}$ be the set of previously matched couples of relationships. For each node matching $<u,u'> \in V_{matched}$, where $u \in V_{Q}$ and $u' \in V_{G}$, BB-Graph applies the \textit{reciprocal node branching process}, that is, the process of expanding partially matched graph piece in order to complete to an exact match by following the adjacent relationships of $u$ and $u'$ simultaneously. In order to decide which node matching is used in reciprocal node branching process at that moment, all the node matchings are kept in a stack which we denote by $S$. When BB-Graph pops a node matching $<u,u'>$ from $S$, it applies the reciprocal node branching as follows: Firstly, it detects the non-matched relationships adjacent to query node $u$. Next, for each of those relationships, it tries to find candidates among the relationships adjacent to $u'$. While determining the candidates, it checks whether the \textit{matching relationship principal} (previously mentioned in this Section) holds or not. After the filtering, there may appear more than one candidate for a relationship. BB-Graph collects all the candidates in a separate list for each non-matched relationship adjacent to $u$, as specified in Algorithm \ref{alg:branch_nodes}, lines 2-4. A case that there is no candidate for a relationship never occurs, because we always match a query node $u$ with a database node $u'$ provided that $u'$ has a degree at least as $u$ has for each different group of relationships.

\begin{algorithm}[ht]
	\SetKwInput{Input}{Input}
	\SetKwInput{Output}{Output}
	\LinesNumbered
	\DontPrintSemicolon
	\BlankLine
	\Input{global variables $S$, $V_{matched}$, $E_{matched}$ \; \qquad \quad 
		  $<u, u'>$: The node match currently being branched}
	\Output{It affects the global variables $S$, $V_{matched}$, $E_{matched}$}
	
	\Begin
	{
		\ForEach{\upshape{non-matched relationship $r_{i}$ adjacent to $u$}}
		{
			$C_{r_{i}} := \{r'_{i} \; | \; r'_{i} \text{ is adjacent to } u' \text{ and } \; r'_{i} \mrp r_{i} \} $
		}
		$k := $ number of non-matched relationships adj. to $u$ \; 
		\textsc{MatchRelationship($1$)} \; 
	}
	\textbf{void} \textsc{MatchRelationship($i$)} \;
	\Begin
	{
		\ForEach{$r'_{i} \in C_{r_{i}}$, $r'_{i}$ \upshape{is not matched}}
		{
			\If{ \textsc{Check($<r_{i}, r'_{i}>$, $<u, u'>$)} }
			{
				Add $<r_{i}, r'_{i}>$ into $E_{matched}$ \;
				$S^{*} := S$, $V^{*}_{matched} := V_{matched}$ \;
				\If{$i \le k$}
				{
					\textsc{MatchRelationship($i$++)}
				}
				\Else
				{
					\textsc{Search}()
				}
				Remove $<r_{i}, r'_{i}>$ from $E_{matched}$ \tcp{Backtracking}
				$S := S^{*}$, $V_{matched} := V^{*}_{matched}$
			}
		}
	}
	\caption{\textsc{BranchNodes}}
	\label{alg:branch_nodes}
\end{algorithm}

Assume that $r_{1}$, $r_{2}$, ...., and $r_{k}$ are the non-matched relationships adjacent to $u$ and $C_{r_{1}}$, $C_{r_{2}}$, .... and $C_{r_{k}}$ are the lists which contain corresponding candidate relationships adjacent to $u'$. In the next stage, BB-Graph picks a candidate relationship $r'_{i} \in C_{r_{i}}$ to match with $r_{i}$ iteratively for each $i$. At each step of iteration, it is checked whether the prospective match $<r_{i},r'_{i}>$ causes a conflict related with the relationship end points, or not. It can easily be seen that if two relationships are matched with each other, then their corresponding end points must also be matched with each other. Here, it is already known that $u$ and $u'$ are matched before. Thus, BB-Graph has to check whether the other end points of $r_{i}$ and $r'_{i}$, say $\nu$ and $\nu'$, match, as shown in Algorithm \ref{alg:check}. It can be one of the following three cases: First, if $\nu$ and $\nu'$ are already matched with each other, then there is no conflict. Second, if $\nu$ is matched with some node other than $\nu'$, or vice versa, it is reported as a conflict. Third, it may be the case that both $\nu$ and $\nu'$ are not-yet-matched nodes. At this point, it is checked whether the nodes satisfy the \textit{matching node principal}, or not. If they satisfy, then they are matched with each other and the new match $<v,v'>$ is added into $V_{matched}$ and pushed into $S$ to apply \textit{reciprocal node branching process} for them later. Otherwise, it is reported as a conflict. As a result, we guarantee that there does not occur an empty candidate list for any relationship in candidate filtering part. After checking the end points of relationships, if there does not exist any reported conflict, then the match $<r_{i}, r'_{i}>$ is approved and added into $E_{matched}$, and BB-Graph goes on the relationship matching procedure from the next iteration for $r_{i+1}$. During the checks, if there is a reported conflict, then the next candidate in $C_{r_{i}}$ is picked and the end point checks are repeated for the prospective matching of $r_{i}$ with its new candidate this time. Nevertheless, it may happen that all the candidate relationships for $r_{i}$ are tried but resulted in a problem. If such a case occurs, then it is understood that there are wrong decisions in the previous matchings. At this point, BB-Graph backtracks to the former relationship matching, say $<r_{i-1}, r'_{i-1}>$, and until getting a non-conflicting match, it tries the next candidates for $r_{i-1}$ this time. If one is obtained, then BB-Graph continues to its normal schedule from the next stage as usual. Otherwise, it backtracks again and applies the same procedure, and goes on in this way, as in Algorithm \ref{alg:branch_nodes}, lines 9-25. 

When BB-Graph backtracks from stage $i$ to $i-1$, all the global data structures are returned to their old version at stage $i-1$; that is, the last relationship and node matchings are cancelled and removed from $V_{matched}$ and $E_{matched}$. Also, $S$ is stored with its old content by doing the opposites of push and pop operations done at stage $i$, as it is given in Algorithm \ref{alg:branch_nodes}, lines 21-22. 

Since our aim is to find all exact matches of the query graph, backtracking is needed not only when a contradiction occurs, but it is also necessary to try all candidate options for each query graph node and relationship with all possible combinations. For that reason, when an isomorphism of the query is found, instead of terminating or restarting, BB-Graph continues its schedule with backtracking. 

When BB-Graph completes the iterative relationship matching part successfully, all the relationships adjacent to $u$ are matched with some relationship adjacent to $u'$. This finishes the reciprocal branching process of two nodes $u$ and $u'$. The rest of algorithm maintains the isomorphism search by recursively repeating the \textit{reciprocal node branching process} for the newly obtained node matchings, as in Algorithm \ref{alg:BB-Graph}, lines 14-29.

In order to illustrate how BB-Graph works, a sample running schedule is given in Figure \ref{fig:sample_execution}. The whole recursive process of reciprocal node branching, relationship matching and backtracking are shown step-by-step for the sample query and the graph database given in Figure \ref{fig:sample_query_database}.

\input{sample_execution.tex}

\subsection{Complexity of The BB-Graph Algorithm}
\label{subsec:complexity}

Here, we give running time complexity (computational complexity) analysis and space complexity analysis of BB-Graph for the worst case scenarios. In the analyses, we denote the number of nodes (vertices) in query graph $Q$ by $|V_{Q}|$, number of nodes (vertices) in graph database $G$ by $|V_{G}|$ and number of relationships (edges) in query graph $Q$ by $|E_{Q}|$. Also, we use $deg_{Q}^{max}$ and $deg_{G}^{max}$ to denote maximum node degree in query graph and maximum node degree in database graph, respectively.

\;\;
\textit{Time Complexity:}
\label{title:time.complexity}
\;

BB-Graph starts with the filtering part. It finds candidates for the starting node by filtering the database nodes depending on the \textit{matching node principal} given in  Section \ref{def:matchingnodeprincipal}. In other words, in order to find all the candidates, label and degree properties of each database node must be checked. Since we work in Neo4j, which holds the information of database nodes grouped by label, the filtering with respect to node labels takes $\mathcal{O}(1)$. In the worst case, all database nodes may have labels of the starting node, which means that there exist $|V_{G}|$ number of candidate nodes that are going to be filtered with respect to their adjacent relationships. Since Neo4j directly gives the information about adjacent reationships of each type for every node, it just remains to ask for this information for every candidate node regarding each relationship of different type adjacent to starting node. Therefore, this part takes $\mathcal{O}(|V_{G}| \times deg_{Q}^{max})$, which also gives the worst case complexity of the total filtering stage. In the end of it, there may be $|V_{G}|$ number of candidates for the starting node in the worst case.

For each candidate of the starting node isomorphism search is conducted. For that reason, the complexity of this part is going to be $|V_{G}| \times \mathcal{O}(\textsc{Search()})$. Because the function \textsc{Search()} and the function \textsc{BranchNodes()} work in a mutual recursive manner, the cost of \textsc{Search()} actually equals to the cost of \textsc{BranchNodes()}. To be able to calculate the complexity of \textsc{BranchNodes}, we should analyse the computational cost of the operations done in one recursive depth of the branching procedure.

\begin{algorithm}[h]
	\SetKwInput{Input}{Input}
	\SetKwInput{Output}{Output}
	\LinesNumbered
	\DontPrintSemicolon
	\BlankLine
	\Input{global variables $S$, $V_{matched}$, $E_{matched}$ \; \qquad \quad
		   $<r_{i}, r'_{i}>$: The relationship match to be checked \; \qquad \quad
		   $<u, u'>$: The node match currently being branched}
	\Output{Boolean, depending on the existence of any contradictory situation}
	
	\Begin
	{
		$\nu_{i} :=$ The end point of $r_{i}$ other than $u_{i}$ \;
		$\nu'_{i} :=$ The end point of $r'_{i}$ other than $u'_{i}$ \;
		\If{$\exists v_{x} \neq \nu_{i}$ \upshape{ s.t. } $<\nu_{x}, \nu'_{i}> \in V_{matched}$}
		{
			return \space $false$
		}
		\If{$\exists \nu'_{x} \neq \nu'_{i}$ \upshape{ s.t. } $<\nu_{i}, \nu'_{x}> \in V_{matched}$}
		{
			return \space $false$
		}
		\If{$<\nu_{i}, \nu'_{i}> \notin V_{matched}$}
		{
			\If{$\nu_{i} \mnp \nu'_{i}$}
			{
				Push $<\nu_{i}, \nu'_{i}>$ into $S$ \;
				Add $<\nu_{i}, \nu'_{i}>$ into $V_{matched}$
			}
			\Else
			{
				return \space $false$
			}
		}
		return \space $true$
	}
	\caption{\textsc{Check}}
	\label{alg:check}
\end{algorithm}

In the branching procedure (Algorithm \ref{alg:branch_nodes}), initially, all non-matched relationships of the current query node are detected and for each of them candidate relationship sets are constructed. If we assume that there are $x_{i}$ number of non-matched relationships for the current node in the depth $i$ of the recursive process, complexity of candidate relationship set construction becomes $\mathcal{O}(x_{i})$. For each non-matched relationship, there can be at most $deg_{G}^{max}$ number of candidate relationships (If there are more than one different type of non-matched relationships adjacent to the current node, then it is certain that the candidate set size is less than $deg_{G}^{max}$. Nevertheless, we take $deg_{G}^{max}$ as an upper bound for the candidate set size of each non-matched query relationship). This means, there occurs at most $(deg_{G}^{max})^{(x_{i})}$ number of different combinations of relationship matching for a query node. Since a query node can have at most $deg_{Q}^{max}$ non-matched relationships, the upper bound for a number of combinations of relationship matching becomes $(deg_{G}^{max})^{(deg_{Q}^{max})}$. At each candidate relationship selection, it is checked whether there occurs any conflict with the match through \textsc{Check()} function (Algorithm \ref{alg:check}). In \textsc{Check()}, it is checked whether the end nodes of both relationships given by the function argument exist among the already matched query and database nodes. Since we use hash tables to understand which nodes are matched, the only thing that costs in this function is checking \textit{matching node principle} between nodes (line 10). For that reason, complexity of this function becomes $\mathcal{O}(deg_{Q}^{max})$. Since \textsc{Check} function is called $x_{i}$, namely $deg_{Q}^{max}$ at most, many times for each combination of relationship matching, at each depth of recursive branching process the cost becomes $\mathcal{O}( (deg_{G}^{max})^{(deg_{Q}^{max})} \times deg_{Q}^{max} \times deg_{Q}^{max})$. Since the maximum depth of the recursive branching can be equal to the number of query nodes, $|V_{Q}|$, the overall complexity of the branching procedure becomes 
$\mathcal{O}( |V_{Q}| \times (deg_{G}^{max})^{(deg_{Q}^{max})} \times (deg_{Q}^{max})^{2} )$.

To conclude, the computational cost of BB-Graph equals to the summation of the cost of the filtering part and the cost of the searching part, which is, $\mathcal{O}(|V_{G}| \times deg_{Q}^{max}) + \mathcal{O}( |V_{G}| \times |V_{Q}| \times (deg_{G}^{max})^{(deg_{Q}^{max})} \times (deg_{Q}^{max})^{2})$. However, the cost of the filtering part is negligible when it is compared to the cost of the searching part. As a result, the running time complexity of BB-Graph becomes \boldmath{$\mathcal{O}( |V_{G}| \times |V_{Q}| \times (deg_{G}^{max})^{(deg_{Q}^{max})} \times (deg_{Q}^{max})^{2})$}\unboldmath.

\;\;
\textit{Space Complexity:}
\label{title:space.complexity}
\;

For the filtering part, BB-Graph uses a list to hold the candidate database nodes for the starting node. Since there can be maximum $|V_{G}|$ number of candidates for the starting node (that is, all the database nodes), there is a need for a list of size $|V_{G}|$ in the worst case. Also, for each query relationship, there is a constructed set of candidate relationships. Since there can be maximum $deg_{G}^{max}$ number of candidates for a query relationship, size of a set of candidate relationships becomes $deg_{G}^{max}$. The case in which the recursive computation reaches to the deepest value is the worst case requiring the largest storage for the sets of candidate relationships. Since all of the query relationships are in the process in the highest depth of the recursion, sets of candidate relationships consume $|E_{Q}| \times deg_{G}^{max}$ units of space at most. When the algorithm backtracks to the previous depth of recursion, the space reserved for containing sets of candidate relationships constructed at that depth is released. Therefore $|E_{Q}| \times deg_{G}^{max}$ is the maximum value of the storage needed by them. Additionally, there exists global variables consuming some memory. There are two global hashing maps used to hold already matched query graph and database graph items; one maps for already matched nodes and the other maps for already matched relationships. These two maps include information of which query node or relationship is matched with which database node or relationship, respectively. For that reason, they can consume at most $2 \times |V_{Q}|$ and $ 2 \times |E_{Q}|$ units of storage, respectively. Lastly, the stack which is used to hold node matchings, which are not yet sent to \textit{reciprocal node branching process} needs $2 \times |V_{Q}|$ units of storage, that is, the space required for the case that starting node is adjacent to all the remaining query nodes. Throughout the whole execution of BB-Graph, all the other variables consume quite little space and can be neglected. Consequently, the space complexity of BB-Graph equals to summation of all the mentioned parameters appeared in the worst case, which is \boldmath{$\mathcal{O}(|V_{G}| + |V_{Q}| + |E_{Q}| \times deg_{G}^{max})$}\unboldmath.

% EXPERIMENTAL WORK

\section{EXPERIMENTS and EVALUATION}
\label{sec:experimental.work}

We compare the performances of BB-Graph, Cypher and GraphQL on Population Database and Bank Database which both are provided by Kale Yaz{\i}l{\i}m, which consist of directed big graphs with 70 millions of nodes and 100 thousands of nodes, respectively. In addition, we also use WorldCup Database, which is publicly available by Neo4j Team, which consists of a directed graph with 45 thousands of nodes. In Table \ref{table:databaseFeatures}, features of Population, Bank and WorldCup Databases are given in detail. As can be noticed, Population Database is much bigger than the ones used in \cite{lee2012depth}, as shown in Table \ref{table:databaseFeatures}.

For the experiments, we use 10 real-world reasonably complex queries for Population Database and 5 real-world queries for each of WorldCup and Bank Databases where each query has different number of nodes and relationships and also has different types of node labels and relationships. For each query, BB-Graph and Cypher experiments have been repeated 10 times, and the averages of the elapsed times are used. For all GraphQL experiments refinement-level is adjusted to one (1) and the refining process has been repeated as many times as the number of nodes in a query graph.  In the experiments of these three algorithms, all the exact matches are found in a continuous time interval without any break. The total elapsed time is calculated in order to compare the performances of the algorithms.  

\subsection{Setup}
\label{subsec:setup}
All the experiments were conducted on the same server with Intel Octa Core 2.27GHz, 8 GB of main memory, and 100 GB hard disk, running Debian GNU, Linux 7.8 (wheezy). The Neo4j version used is 2.3.1. BB-Graph and GraphQL were implemented in Java on Eclipse and Cypher queries were called in Java.

\begin{table} [H]
	\begin{center}
		\resizebox{1\textwidth}{!}{
			\begin{tabular}{ | l | l | l | p{1.8cm} |}
				\hline
				& \textbf{WorldCup DB} & \textbf{Bank DB} & \textbf{Population DB} \\ \hline
				Size & 20 MiB & 510.23 MiB & 10.32 GiB \\ \hline
				\# of graphs & 1 & 1 & 1 \\ \hline
				\# of nodes & 45348 & 105085 & 70422787 \\ \hline
				\# of relationships & 86577 & 107898 & 77163109 \\ \hline
				\# of distinct node labels & 12 & 15 & 14 \\ \hline
				\# of distinct relationship types & 17 & 18 & 18 \\ \hline
				Avg. \# of labels per node & 1 & 1 & 1 \\ \hline
			\end{tabular}
		}
		\caption{General Features of Population, Bank and WorldCup Databases}
		\label{table:databaseFeatures}
	\end{center}
\end{table}

\subsection{Experimental Analysis}

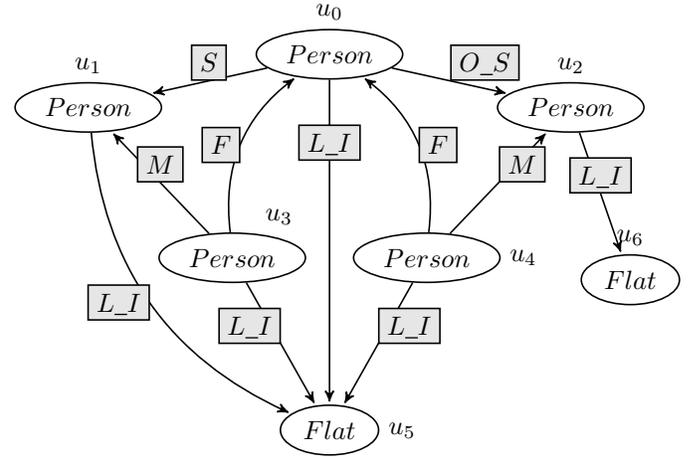
\begin{figure} [!ht]
	\centering
 	\begin{tikzpicture}
 		[->,>=stealth',shorten >=1pt,auto,node distance=1cm, semithick]
 		\tikzset{vertex/.style = {shape=ellipse,draw,minimum size=1.5em}}
 		\tikzset{edge/.style = {->}}
 		
 		\node[vertex, label=$u_{0}$]								(0) at (0,0)		{$Person$};
 		\node[vertex, label=$u_{1}$, below left of=0]						(1) at (-2.5,0)		{$Person$};
 		\node[vertex, label=$u_{2}$, below right of=0]					(2)	at (2.5,0)		{$Person$};
 		\node[vertex, label=above right:$u_{3}$, below right of=1]	(3) at (-2,-2)		{$Person$};
 		\node[vertex, label=right:$u_{4}$, below left of=2]			(4)	at (2,-2)		{$Person$};
 		\node[vertex, label=right:$u_{5}$, below of=0]				(5)	at (0,-4)		{$Flat$};
 		\node[vertex, draw=none, below of=0]				(7)	at (0,-4.2)		{};
 		\node[vertex, label=$u_{6}$, below of=2]				(6)	at (4,-2)		{$Flat$};
 		
 		\path
 		(0)	edge 				node[above, draw, fill=gray!20]					{$S$}		(1)
 		edge 				node[draw, fill=gray!20]						{$O\_S$}	(2)
 		edge 				node[above, near start, draw, fill=gray!20]		{$L\_I$}	(5)
 		(1)	edge [bend right]	node[left, draw, fill=gray!20]					{$L\_I$}	(5)
 		(2)	edge 				node[above, draw, fill=gray!20]					{$L\_I$}	(6)
 		(3)	edge [bend left]	node[left, draw, fill=gray!20]					{$F$}		(0)
 		edge 				node[above, draw, fill=gray!20]					{$M$}		(1)
 		edge 				node[above left, draw, fill=gray!20]			{$L\_I$}	(5)
 		(4)	edge [bend right]	node[right, draw, fill=gray!20]					{$F$}		(0)
 		edge 				node[above right, draw, fill=gray!20]			{$M$}		(2)
 		edge 				node[above right, draw, fill=gray!20]			{$L\_I$}	(5);
 		
 	\end{tikzpicture}
	\caption{An example for a complex query: \textit{"Find the families consisting of a man, a wife, at least two children, one from an ex-wife, who live at the same address in Population Database"}} 
	\label{fig:complex.query}
\end{figure}

\begin{figure} [!h]
	\centering
		\begin{tikzpicture}
			[->,>=stealth',shorten >=1pt,auto,node distance=1.5cm, semithick]
			\tikzset{vertex/.style = {shape=ellipse,draw,minimum size=1.5em}}
			\tikzset{edge/.style = {->}}
			
			\node[vertex, label=$u_{0}$]					(0) at (0,0)	{$Country$};
			\node[vertex, label=$u_{1}$, below right of=0, xshift=50]		(1)	{$Squad$};
			\node[vertex, label=$u_{2}$, below left of=1, xshift=-50]		(2)	{$Player$};
			\node[vertex, label=$u_{3}$, below right of=2, xshift=50]		(3)	{$Squad$};
			\node[vertex, label=$u_{4}$, below left of=3, xshift=-50]		(4)	{$Country$};
			\node[vertex, draw=none, below of=4, yshift=25]							(5) {};
			
			\path
			(0)	edge 	node[draw, fill=gray!20]	{$N\_S$}	(1)
			(2)	edge	node[draw, fill=gray!20]	{$I\_S$}	(1)
				edge	node[draw, fill=gray!20]	{$I\_S$}	(3)
			(4) edge	node[draw, fill=gray!20] 	{$N\_S$}	(3);
								
 	\end{tikzpicture}
	\caption{An example for a path query: \textit{"Find the players who join the squad of different countries in WorldCup Database"}} 
	\label{fig:path.query}
\end{figure}
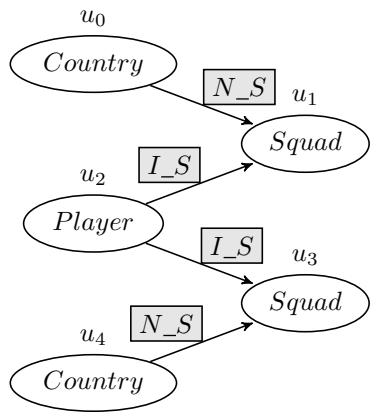

\begin{figure} [!h]
	\centering
	\resizebox{1\textwidth}{!}{
	\begin{tikzpicture}
	[->,>=stealth',shorten >=1pt,auto,node distance=2cm, semithick]
	\tikzset{vertex/.style = {shape=ellipse,draw,minimum size=1.5em}}
	\tikzset{edge/.style = {->}}
	
	\node[vertex, label=$u_{0}$]								(0) at (0,0)		{$Player$};
	\node[vertex, label=$u_{1}$, below left of=0]				(1) at (-2.5,-0.5)	{$Performance$};
	\node[vertex, label=below right:$u_{2}$, below of=0]		(2)	at (0,-0.5)		{$Performance$};
	\node[vertex, label=$u_{3}$, below right of=0]				(3) at (2.5,-0.5)	{$Performance$};
	\node[vertex, label=$u_{4}$, below left of=2]				(4)	at (-2.5,-3)	{$Match$};
	\node[vertex, label=right:$u_{5}$, below of=2]				(5) at (0,-3)		{$Match$};
	\node[vertex, label=$u_{6}$, below right of=2]				(6) at (2.5,-3)		{$Match$};
	\node[vertex, label=above left:$u_{7}$, below left of=5]	(7)	at (-3,-6)		{$WorldCup$};
	\node[vertex, label=above right:$u_{8}$, below of=5]		(8)	at (0,-6)		{$WorldCup$};
	\node[vertex, label=above right:$u_{9}$, below right of=5]	(9)	at (3,-6)		{$WorldCup$};
	
	\path
	(0)	edge 	node[above, draw, fill=gray!20]		{$ST$}		(1)
		edge 	node[above, draw, fill=gray!20]		{$ST$}		(2)
		edge 	node[above, draw, fill=gray!20]		{$ST$}		(3)
	(1)	edge 	node[above, draw, fill=gray!20]		{$I\_M$}	(4)
	(2)	edge 	node[left, draw, fill=gray!20]		{$I\_M$}	(5)
	(3)	edge 	node[above, draw, fill=gray!20]		{$I\_M$}	(6)
	(4)	edge 	node[above, draw, fill=gray!20]		{$C\_M$}	(7)
	(5)	edge 	node[above, draw, fill=gray!20]		{$C\_M$}	(8)
	(6)	edge 	node[above, draw, fill=gray!20]		{$C\_M$}	(9);
	
	\end{tikzpicture}}
	
	\caption{An example for a tree query: \textit{"Find the players who participate in at least 3 different World Cup Tournaments in WorldCup Database"}} 
	\label{fig:tree.query}
\end{figure}
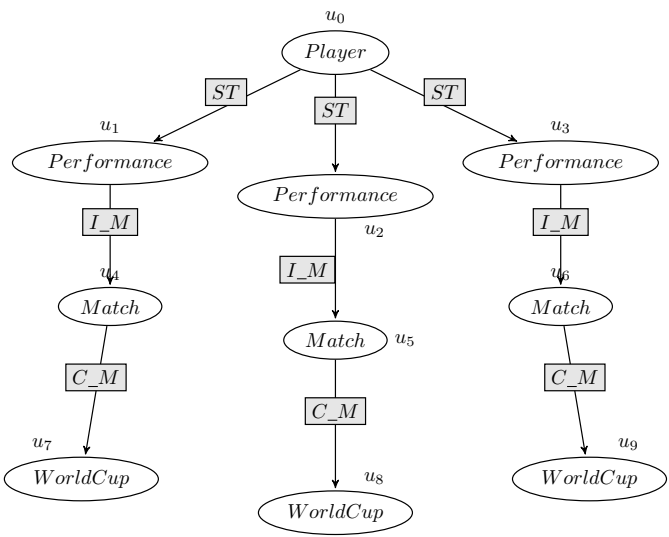

We group the experimental results according to the query type constructed by node matching orders used in BB-Graph. There exist 3 types of queries that we use in our experiments: First, if a query includes cyclic paths on it, we call that as a \textit{complex query}. Second, if a query itself is a single path and BB-Graph starts matching from one of its end nodes, we simply call that as a \textit{path query}. Third, if a query does not contain any cycles and BB-Graph matches its nodes in a tree manner, we call that as a \textit{tree query}. Figures \ref{fig:complex.query}, \ref{fig:path.query} and \ref{fig:tree.query} show a sample complex query, a sample path query and a sample tree query, respectively.

\;\;\;\;
\textit{Complex Queries:}
\label{title:complex.queries}

Table \ref{table:complex_queries} shows the experimental results obtained for distinct complex-type query graphs executed over Worldcup, Bank and Population Databases.

\begin{table*} [!ht]
	\begin{center}
		\resizebox{1\textwidth}{!}{
	    \begin{tabular}{ | c | c | c | c | c | c | c | p{2.3cm} |}
		    \hline
		    \textbf{Query} & \textbf{Database} 
		    & \textbf{\pbox{50cm}{Number \\ of \\ Nodes}} 
		    & \textbf{\pbox{50cm}{Number \\ of \\ Relationships}} 
		    & \textbf{\pbox{50cm}{Number \\ of \\ Cycles}}
		    & \textbf{\pbox{50cm}{Peformance \\ of \\ GraphQL (sec.)}} 
		    & \textbf{\pbox{50cm}{Performance \\ of \\ Cypher (sec.)}}
		    & \textbf{\pbox{50cm}{Performance \\ of \\ BB-Graph (sec.)}} \\ \hline
		    
			\textbf{Q-1} & WorldCup & 4 & 4 & 1 & 1112.5 & 10.1 & 1.9 \\ \hline
			\textbf{Q-2} & WorldCup & 7 & 8 & 2 & 17.9 & 7.8 & 3.9 \\ \hline
			\textbf{Q-3} & WorldCup & 10 & 12 & 3 & 8.1 & 26.7 & 7.9 \\ \hline
			\textbf{Q-4} & Bank & 4 & 4 & 1 & 238.9 & 1.5 & 1.4 \\ \hline
			\textbf{Q-5} & Bank & 5 & 5 & 1 & 339.0 & 2.1 & 2.0 \\ \hline
			\textbf{Q-6} & Bank & 6 & 6 & 1 & 630.8 & 3.6 & 2.7 \\ \hline
			\textbf{Q-7} & Bank & 9 & 9 & 1 & $>$ 1800.0 & 8.9 & 6.0 \\ \hline
			\textbf{Q-8} & Bank & 12 & 12 & 1 & $>$ 1800.0 & 226.8 & 16.0 \\ \hline
			\textbf{Q-9} & Population & 6 & 12 & 7 & $>$ 1800.0 & 189.9 & 138.1 \\ \hline
			\textbf{Q-10} & Population & 5 & 7 & 4 & $>$ 1800.0 & 24.2 & 22.3 \\ \hline
			\textbf{Q-11} & Population & 5 & 5 & 1 & $>$ 1800.0 & 14.8 & 17.3 \\ \hline
		    \textbf{Q-12} & Population & 6 & 6 & 1 & $>$ 1800.0 & 92.4 & 54.5 \\ \hline
		    \textbf{Q-13} & Population & 7 & 11 & 5 & $>$ 1800.0 & 14.4 & 8.5 \\ \hline
		   	\textbf{Q-14} & Population & 7 & 11 & 4 & $>$ 1800.0 & 13.7 & 14.0 \\ \hline
		   	\textbf{Q-15} & Population & 13 & 25 & 13 & $>$ 1800.0 & $>$ 1800.0 & 502.7 \\ \hline
		   	\textbf{Q-16} & Population & 10 & 14 & 3 & $>$ 1800.0 & 226.7 & 73.4 \\ \hline
		   
	    \end{tabular}}
	    \caption{Query Response Times for Complex Queries}
	    \label{table:complex_queries}
	\end{center}
\end{table*}

As it can be clearly seen from Table \ref{table:complex_queries}, BB-Graph performs so much better than both Cypher and GraphQL for a large majority of complex-type queries. Generally, BB-Graph has higher performance than Cypher for the queries that we have tested, except the queries Q-11 and Q-14, for which it performs as nearly good as Cypher. 

For the complex queries executed in WorldCup Database which are Q-1, Q-2 and Q-3 consisting of 4 nodes and 4 relationships with 1 cycle, 7 nodes and 8 relationships with 2 cycles and 10 nodes and 12 relationships with 3 cycles, respectively, BB-Graph has the best performance among all the algorithms. Although GraphQL has the worst performance for Q-1 and Q-2, it achieves a very close performance to BB-Graph for Q-3. We think that it is probably caused by the fact that the existence of a very unique node label in Q-3 made the number of candidate nodes very small. Therefore, the entire search does not take a long time. 

In Bank Database, despite the fact that GraphQL succeeds in returning the results for Q-4 consisting of 4 nodes and 4 relationships with 1 cycle, Q-5 consisting of 5 nodes and 5 relationships with 1 cycle and Q-6 consisting of 6 nodes and 6 relationships with 1 cycle, its performance clearly falls behind the performances of Cypher and BB-Graph. On the other hand, for the other two queries in Bank Database, which are Q-7 consisting of 9 nodes and 9 relationships with 1 cycle and Q-8 consisting of 12 nodes and 12 relationships with 1 cycle, GraphQL could not terminate in 30 minutes. Lastly, for all of the five queries run in Bank Database, as it can be seen from Table \ref{table:complex_queries}, BB-Graph performs better than Cypher.

 From those queries executed in Population Database, for Q-9, Q-10, Q-12, Q-13, Q-15 and Q-16 consisting of 6 nodes and 12 relationships with 7 cycles, 5 nodes and 7 relationships with 4 cycles, 6 nodes and 6 relationships with 1 cycle, 7 nodes and 11 relationships with 5 cycles, 13 nodes and 25 relationships with 13 cycles and 10 nodes and 14 relationships with 3 cycles, respectively, BB-Graph performs much better than Cypher. For Q-15, which is the most complex and the largest query among all in the experiments, Cypher could not finish its execution in a reasonable time. Interestingly, although Q-13 and Q-14 have the same number of nodes, relationships and even the node labels, the performance of BB-Graph changes. The reason for that as follows: While both of those queries have the same number of items, they have different structuring such as the neighbourhoods, node degrees and the number of cycles, which also specialize a graph. For that reason, performance of BB-Graph changes with the number of false candidates that depends on how much a query graph is specialized with its characteristics. Lastly, in Population Database, GraphQL could not terminate in 30 minutes for any of complex queries executed.

\;\;\;\;
\textit{Path Queries:}
\label{title:path.queries}
\;\;\;\;

Table \ref{table:path_queries} shows the experimental results obtained for distinct path-type query graphs executed over Worldcup and Population Databases.

\begin{table*} [!ht]
	\begin{center}
		\resizebox{1\textwidth}{!}{
	    \begin{tabular}{ | c | c | c | c | c | c | c | p{6cm} |}
		    \hline
		    \textbf{Query} & \textbf{Database} 
		    & \textbf{\pbox{50cm}{Number of \\ Nodes}} 
		    & \textbf{\pbox{50cm}{Number of \\ Relationships}} 
		    & \textbf{\pbox{50cm}{Peformance of \\ GraphQL (sec.)}} 
		    & \textbf{\pbox{50cm}{Performance of \\ Cypher (sec.)}}
		    & \textbf{\pbox{50cm}{Performance of \\ BB-Graph (sec.)}} \\ \hline
		    
		    \textbf{Q-17} & WorldCup & 5 & 4 & 28.5 & 1.6 & 0.4 \\ \hline
		    \textbf{Q-18} & Population & 8 & 7 & $>$ 1800.0 & 26.2 & 2.2 \\ \hline
		    \textbf{Q-19} & Population & 4 & 3 & $>$ 1800.0 & 12.5 & 18.9 \\ \hline
		 
	    \end{tabular}}
	    \caption{Query Response Times for Path Queries}
	    \label{table:path_queries}
	\end{center}
\end{table*}

From Table \ref{table:path_queries}, it can be seen that BB-Graph has the best performance for 2 path-type queries out of 3. The first path query, Q-17, consists of 5 nodes and 4 relationships executed over WorldCup Database. For this query, BB-Graph shows 4 times better performance than Cypher. For Q-17, GraphQL shows rather low performance when compared to BB-Graph's and Cypher's performances. On the other hand, this is the only path query that GraphQL could complete its execution in 30 minutes. Since Population Database is much more complex, dense and bigger than WorldCup Database, GraphQL exceeds our reasonable time limit, which is 30 minutes, for the other two path queries. The second path query, Q-18, consists of 8 nodes and 7 relationships executed over Population Database. This query is a recursive query, with the depth of 4. For Q-18, again BB-Graph performs much better than Cypher. On the other hand, for Q-19, which is the third path query consisting of 4 nodes and 3 relationships, Cypher peforms almost the same as BB-Graph.

\textit{Tree Queries:}
\label{title:tree.queries}

Table \ref{table:tree_queries} shows the experimental results obtained for distinct tree-type query graphs executed over Worldcup and Population Databases.

\begin{table*} [!ht]
	\begin{center}
		\resizebox{1\textwidth}{!}{
	    \begin{tabular}{ | c | c | c | c | c | c | c | p{6cm} |}
		    \hline
		    \textbf{Query} & \textbf{Database} 
		    & \textbf{\pbox{50cm}{Number of \\ Nodes}} 
		    & \textbf{\pbox{50cm}{Number of \\ Relationships}} 
		    & \textbf{\pbox{50cm}{Peformance of \\ GraphQL (sec.)}} 
		    & \textbf{\pbox{50cm}{Performance of \\ Cypher (sec.)}}
		    & \textbf{\pbox{50cm}{Performance of \\ BB-Graph (sec.)}} \\ \hline
		    
		    \textbf{Q-20} & WorldCup & 10 & 9 & $>$ 1800.0 & 26.9 & 35.7 \\ \hline
		    \textbf{Q-21} & Population & 4 & 3 & $>$ 1800.0 & 12.5 & 24.1 \\ \hline
		 
	    \end{tabular}}
	    \caption{Query Response Times for Tree Queries}
	    \label{table:tree_queries}
	\end{center}
\end{table*}

We have executed 2 tree-type queries; one in WorldCup Database and the
other in Population Database. The first tree query, Q-20, consists of 10 nodes
and 9 relationships on which there exist 3 identical paths including 3 nodes such that all the 3 paths are connected to a common node and construct a tree with 3 branches. Similarly, the second query, Q-21, consists of 4 nodes and 3 relationships on which there exist 2 identical paths including 2 nodes such that the paths are connected to each other with an edge. It can be seen from Table \ref{table:tree_queries} that Cypher has a better performance than BB-Graph for this type of queries. We think that the reason for the Cypher's good performance is that Cypher applies an efficiency rule for the queries which include identical patterns by conducting a search only for one of those patterns instead of searching for each. Since we do not analyze the query graphs in terms of their analytical features which have such short cuts, performance of BB-Graph falls somewhat behind the performance of Cypher in those cases. As a future work, BB-Graph can be enhanced in order to increase the performance for such cases as well.

\;\;\;\;
\textit{Matching Order Analysis For BB-Graph}
\label{title:matching.order}
\;\;\;\;

Table \ref{table:matching_order} shows the effect of different node matching orders for some of the queries executed by BB-Graph in Population Database.

\begin{table*} [!ht]
	\begin{center}
		\resizebox{1\textwidth}{!}{
	    \begin{tabular}{ | c | c | c | c | c | c | p{6cm} |}
		    \hline
		    \textbf{Query} & \textbf{Database} 
		    & \textbf{\pbox{50cm}{Number of \\ Nodes}} 
		    & \textbf{\pbox{50cm}{Number of \\ Relationships}} 
		    & \textbf{\pbox{75cm}{Performance of \\ BB-Graph (sec.) \\ Matching Order-1}}
		    & \textbf{\pbox{75cm}{Performance of \\ BB-Graph (sec.) \\ Matching Order-2}} \\ \hline
		    
		    \textbf{Q-18} & Population & 8 & 7 & 41.4 & 2.2 \\ \hline
		    \textbf{Q-19, Q-21} & Population & 4 & 3 & 24.1 & 18.9 \\ \hline
		    \textbf{Q-11} & Population & 5 & 5 & 33.3 & 17.3 \\ \hline
		    \textbf{Q-12} & Population & 6 & 6 & 101.4 & 54.5 \\ \hline
		 
	    \end{tabular}}
	    \caption{Query Response Times for Different Matching Orders in BB-Graph}
	    \label{table:matching_order}
	\end{center}
\end{table*}

As it is seen from Table \ref{table:matching_order}, when we change the node matching order followed by BB-Graph Algorithm (by just changing the starting node and then the rest is determined by the algorithm itself), the performance considerably changes. For Q-18, which is the path query explained above, the reason of the change in performance is the $N$-1, 1-1 or $N$-$M$ property of relationship types between the node labels. Clearly, for a $N$-1 type of relationship, it is better to match the node in '$N$' side first, and then the node in '1' side next, instead of matching those in the reverse order because the number of candidate nodes automatically  decreases to 1 in the first case whereas there may occur $N$ candidates in the second case. Thus, for Q-18, when we change the path following direction from 1-$N$ to $N$-1, the performance increases almost 20 times. For Q-21, which is a tree query, the performance increases when we match its nodes as a path query that corresponds to Q-19. Additionally, for Q-11 and Q-12, the increase in performance occurs when BB-Graph starts matching from the node that is rarer due to its semantics appearing with its neighbourhood. As a result, since the node matching order may result in a change in the number of false candidates for query nodes, it highly affects the performance of BB-Graph.

% CONCLUSION

\section{CONCLUSION}
\label{sec:conclusion}

In this paper, a new algorithm, BB-Graph, for Subgraph Isomorphism Problem is introduced. BB-Graph uses branch-and-bound technique to match each node and relationship of query graph with its candidates and uses backtracking approach for the other possible candidates. Different from the current algorithms, BB-Graph does not find candidates of each query graph node all across the graph database. After matching the first query node, BB-Graph searches candidates for the other query graph nodes and relationships in local region of the first-matched database node. Our experiments are conducted with different types of queries on two different real-world databases, namely Population Database and WorldCup Database, and one simulated database, namely Bank Database. According to our experimental results, GraphQL is not scalable enough for querying in very big databases like Population Database. Although GraphQL shows similar performance with BB-Graph for one of the queries in WorldCup Database, generally GraphQL performs worse than Cypher and BB-Graph due to its strategy of getting the candidates from the whole database and trying to match even some irrelevant ones. On comparing BB-Graph with Cypher, our experimental results show that BB-Graph performs better than Cypher for most of the query types, for both large and small databases.

According to our experimental results, the performance of BB-Graph algorithm does not depend on just one factor like the number of nodes or relationships. Actually, many other features, such as frequency of query node labels and query relationship types in the database graph, number of cycles in query graph, structure of query graph (whether it is a path or something more complex) and also the semantic design of relationships like $1-N$ or $N-N$, affect the computation time. As a future study, it might be possible to improve the performance of BB-Graph by taking all the query and graph database features mentioned above into consideration.

\ifCLASSOPTIONcompsoc
  \section*{Acknowledgments}
\else
  \section*{Acknowledgment}
\fi

We thank to KALE YAZILIM for sharing their resources, especially the real-world database, and giving the opportunity to work with them.

\ifCLASSOPTIONcaptionsoff
  \newpage
\fi

\bibliographystyle{IEEEtran}
\bibliography{references}


\begin{IEEEbiography}[{\includegraphics[width=1in,height=1.25in,clip,keepaspectratio]{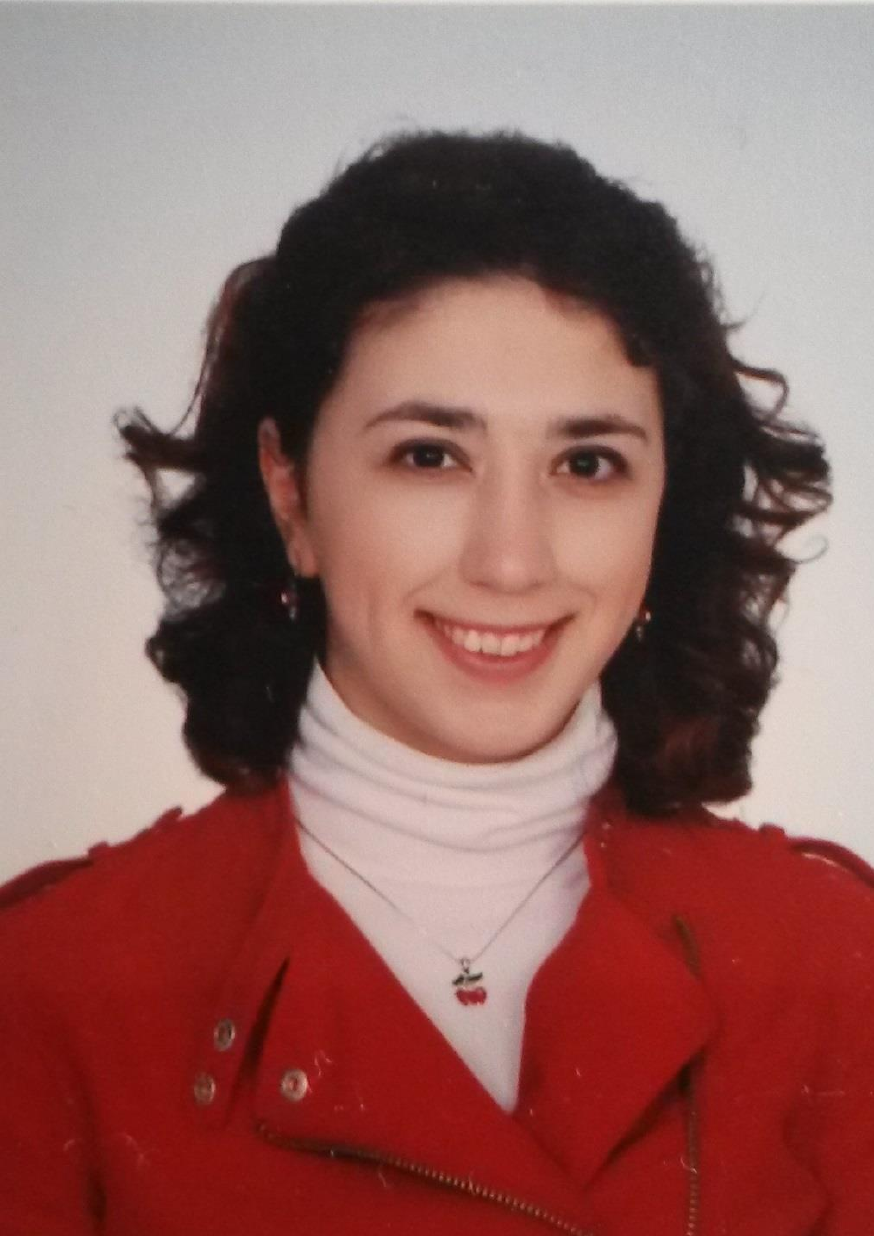}}]{Merve Asiler}
	is a research and teaching assistant and a graduate student pursuing her Ph.D. degree in the Department of Computer Engineering at Middle East Technical University (METU). She received the B.S. degree in the Department of Mathematics at Middle East Technical University in 2013 with a double major in the Department of Computer Engineering, and received the M.Sc. degree from the Department of Computer Engineering, Middle East Technical University, Ankara Turkey, in 2016. Her research interests include graph algorithms, big data and databases.
\end{IEEEbiography}

\begin{IEEEbiography}[{\includegraphics[width=1in,height=1.25in,clip,keepaspectratio]{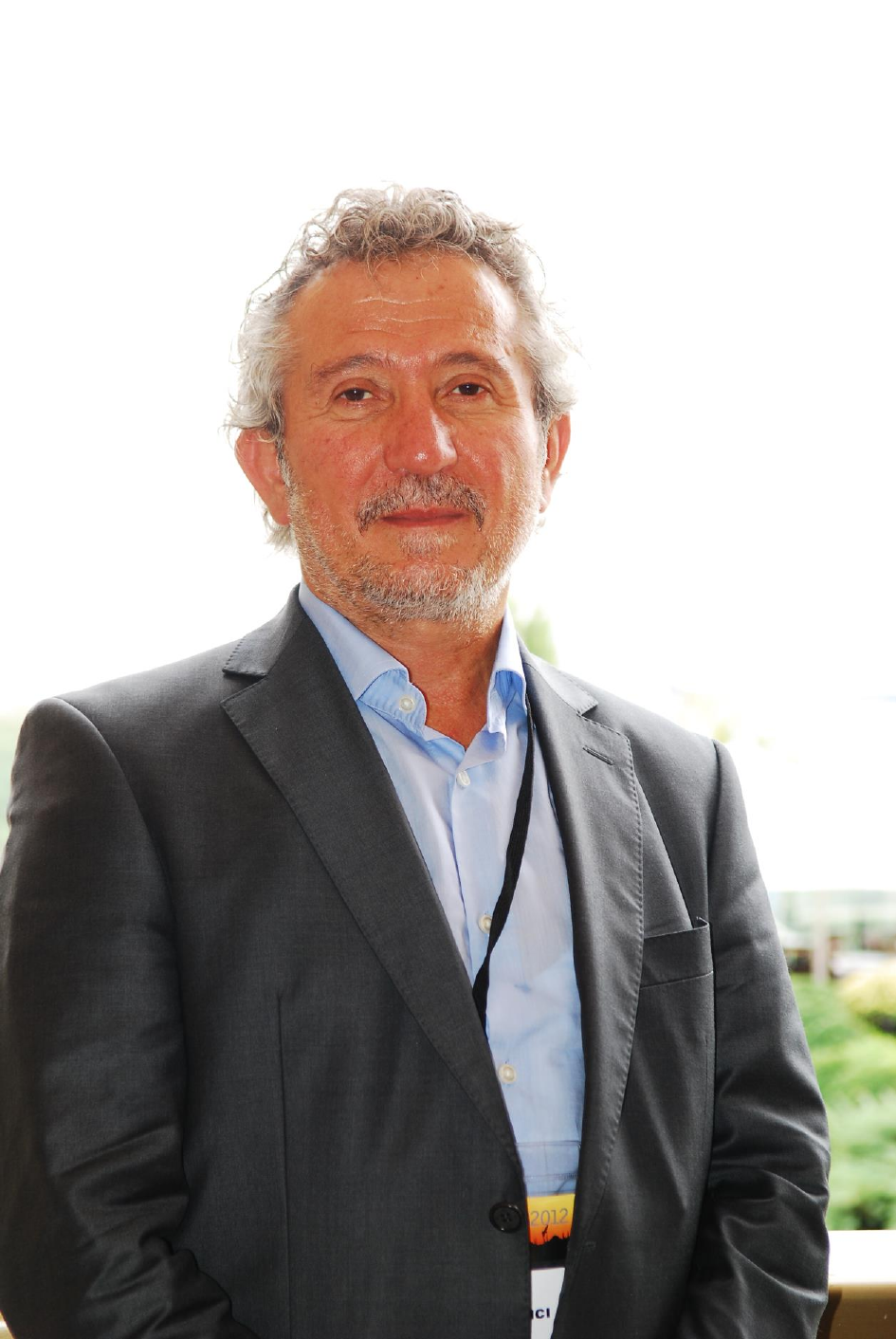}}]{Adnan Yaz{\i}c{\i}}
	received the Ph.D. degree in computer science from the Department of EECS, Tulane University, LA, USA, in 1991. He is a Full professor at Department of Computer Engineering, Middle East Technical University, Ankara, Turkey and currently Chair of the Department of Computer Science at School of Science and Technology in Nazarbayev University in Astana, Kazakhstan. He has published over 200 international technical papers and co-authored/edited three books entitled Fuzzy Database Modeling (Springer), Fuzzy Logic in its 50th Year: New Developments, Directions and Challenges (Springer), and Uncertainty Approaches for Spatial Data Modeling and Processing: A Decision Support Perspective (Springer). His current research interests include intelligent database systems, multimedia and video databases and information retrieval, wireless multimedia sensor networks, data science, and fuzzy database modeling. He was a recipient of the IBM Faculty Award in 2011 and the Parlar Foundations Young Investigator Award in 2001. He was the Conference Co-Chair of the 23rd IEEE International Conference on Data Engineering in 2007, the 38th Very Large Data Bases in 2012, and the 23rd IEEE International Conference on Fuzzy Systems in 2015. He is currently an Associate Editor of the IEEE Transactions on Fuzzy Systems and a member of ACM, the IEEE Computational Intelligence Society and the Fuzzy Systems Technical Committee.
\end{IEEEbiography}

%\begin{IEEEbiographynophoto}{Adnan Yaz{\i}c{\i}}

\end{document}